\documentstyle[a4]{article}
\def\abstract#1{\vskip 7mm 
        \begin{center}{\large Abstract}\par \smallskip
                \begin{minipage}[c]{12cm}
                        \small #1
                \end{minipage}
        \end{center}
}
\def\title#1{\begin{center}{\Large\bf #1}\end{center}}
\def\author#1{\vskip 5mm \begin{center}{#1}\end{center}}
\def\address#1{\begin{center}{\it #1}\end{center}}

\begin{document}

\title{
Entanglement thermodynamics
}

\author{
Shinji Mukohyama$^\dagger$,  
Masafumi Seriu$^\ddagger$
and 
Hideo Kodama$^\dagger$
}

\address{
${}^\dagger$Yukawa Institute for Theoretical Physics \\
Kyoto University, Kyoto 606-01, Japan \\
and \\
${}^\ddagger$ Physics Group, Department of Education \\
Fukui University, Fukui 910, Japan
}

\abstract{
Entanglement entropy is a statistical entropy measuring information loss 
due to coarse-graining corresponding to a spatial division of a system. 
In this paper we construct a thermodynamics (entanglement thermodynamics)
which includes the entanglement entropy as the entropy variable, for a 
massless scalar field in Minkowski, Schwarzschild and 
Reissner-Nordstr{\"o}m spacetimes to understand the statistical origin 
of black-hole thermodynamics. It is shown that the entanglement 
thermodynamics in Minkowski spacetime differs significantly from 
black-hole thermodynamics. On the contrary, the entanglement 
thermodynamics in Schwarzschild and Reissner-Nordstr{\"o}m spacetimes has 
close relevance to black-hole thermodynamics. 
}

\section{Introduction}

In the classical theory of black-hole there appears a thermodynamical 
relation among the parameters describing black-holes. We call it 
black-hole thermodynamics~\cite{Bekenstein}. For example, for a one-parameter 
family of Schwarzschild black-holes parameterized by their mass $M$, the 
same relation as the first law of thermodynamics holds: 
%
\begin{equation}
 dE_{BH} = T_{BH}dS_{BH}\ \ ,	
\end{equation}
where $E_{BH}$, $S_{BH}$ and $T_{BH}$ are defined by
%
\begin{eqnarray}
 E_{BH} & \equiv & M\ \ ,\nonumber\\
 S_{BH} & \equiv & \frac{A_{H}}{4l_{pl}^2}\ \ ,\nonumber\\
 T_{BH} & \equiv & \frac{1}{4\pi R_{H}}\ \ .\label{eqn:EST}
\end{eqnarray}
Here $R_{H}=2Ml_{pl}^2$ and $A_{H}=4\pi R_{H}^2$ are the area radius 
and the area of the event horizon. It is well-known that in general the 
area of the event horizon does not decrease in time classically. Hence 
$S_{BH}$ defined by (\ref{eqn:EST}) does not decrease in time as thermodynamical 
entropy (the second law of black-hole). It can also be shown~\cite{GSL} 
that sum of $S_{BH}$ and entropy of matter field outside the horizon does 
not decrease in time semi-classically (the generalized second law). 
Thus $S_{BH}$ can be regarded as a entropy of a black-hole. Moreover, a 
black-hole with surface gravity $\kappa$ emits thermal radiation at 
the so-called Hawking temperature $\kappa /2\pi$~\cite{Hawking1975}. For 
the Schwarzschild black-hole this Hawking temperature is given by $T_{BH}$. 
From these facts $S_{BH}$ and $T_{BH}$ are called black-hole entropy and 
black hole temperature, respectively.  

There have been many attempts to understand the statistical origin of 
the black-hole entropy~\cite{BH-entropy}. Among them a strong candidate is 
the so-called  entanglement entropy~\cite{BLLS,Srednicki,Sent}. In this 
paper we attempt to construct a thermodynamics which includes 
the entanglement entropy as the entropy. We call it thermodynamics of 
entanglement or entanglement thermodynamics~\cite{paper1}.

\section{Basic ingredients of the entanglement thermodynamics}

First we review the basic concepts of entanglement thermodynamics. 
Let ${\cal F}$ be a Hilbert space constructed from two Hilbert spaces
${\cal F}_{1}$ and ${\cal F}_{2}$ as 
%
\begin{equation}
{\cal F} = {\cal F}_{1} \bar{\otimes} {\cal F}_{2} \ \ \ ,
	\label{eqn:F=F1*F2}
\label{eqn:Hilbert}
\end{equation}
where $\bar{\otimes}$ denotes a tensor product followed by a suitable
completion. From an element $u$ of ${\cal F}$ with unit norm we construct 
an operator $\rho$ (`density operator') by
%
\begin{equation}
 \rho v = (u,v)u \quad {}^\forall v\in{\cal F},
\end{equation}
where $(u,v)$ is the inner product which is antilinear with respect to 
$u$. From $\rho$ we define so-called reduced density operators
$\rho_{1,2}$ by
%
\begin{eqnarray}
 \rho_{1}y & = & \sum_{i,j}e_i(e_i\otimes f_j,\rho (y\otimes f_j))
	\quad {}^\forall y\in{\cal F}_{1}\ \ ,\nonumber\\
 \rho_{2}z & = & \sum_{i,j}f_j(e_i\otimes f_j,\rho (e_i\otimes z))
	\quad {}^\forall z\in{\cal F}_{2}\ \ ,
\end{eqnarray}
where $\{ e_i\}$ and $\{ f_j\}$ are orthonormal bases of ${\cal F}_{1}$ 
and ${\cal F}_{2}$, respectively. The entanglement entropy is defined by 
$S_{ent}=S[\rho_{1}]$ or $S_{ent}=S[\rho_{2}]$ or 
$S_{ent}=S[\rho_{1}\otimes\rho_{2}]$, where 
%
\begin{eqnarray}
 S[\rho_{1,2}] & = & -{\bf Tr}[\rho_{1,2}\ln\rho_{1,2}]\ \ ,
 	\nonumber\\
 S[\rho_{1}\otimes\rho_{2}] & = & S[\rho_{1}]+S[\rho_{1}]\ \ .
\end{eqnarray}
Since it can be shown that $S[\rho_{1}]=S[\rho_{2}]$ in general~\cite{paper1}, 
the three options for $S_{ent}$ are identical up to the factor $2$. For
definition of the entanglement energy we consider the following four
options: 
$E_{ent}=\langle :H_{1}:\rangle$ or $E_{ent}=\langle :H_{2}:\rangle$  
or $E_{ent}=\langle :H_{1}:\rangle +\langle :H_{2}:\rangle$ or 
$E_{ent}=\langle :H_{tot}:\rangle_{\rho_{1}\otimes\rho_{2}}$ , where
%
\begin{eqnarray}
 \langle :H_{1,2}:\rangle & = & {\bf Tr}[\rho :H_{1,2}]\ \ ,
 	\nonumber\\
 \langle :H_{tot}:\rangle_{\rho_{1}\otimes\rho_{2}} & = & 
	{\bf Tr}[(\rho_{1}\otimes\rho_{2}):H_{tot}:]\ \ .
\end{eqnarray}
Here the total hamiltonian $H_{tot}$ is assumed to be decomposed as 
$H_{tot}=H_{1}+H_{2}+H_{int}$, and $:\cdots :$ denotes the normal 
ordering. Finally entanglement temperature $T_{ent}$ is obtained by 
imposing the first law of entanglement thermodynamics:
%
\begin{equation}
 T_{ent}dS_{ent} = dE_{ent}\ \ .
\end{equation}

\section{Discretized theory of a scalar field}

In this paper we consider a massless real scalar field described by 
the action
%
\begin{equation}
 I = -\frac{1}{2}\int dx^4\sqrt{-g}
	\partial^{\mu}\phi \partial_{\mu}\phi\ \ ,
	\label{eqn:action}
\end{equation}
where the background geometry is fixed to be a spherically symmetric 
static spacetime with the metric 
%
\begin{equation}
 ds^2 = -N(\rho)^2 dt^2 + d\rho^2 
 	+r(\rho)^2\left( d\theta + \sin^2{\theta}d\psi^2\right)\ \ .
\end{equation}
For this system we calculate the entanglement entropy and 
the entanglement energy to construct entanglement thermodynamics 
using the methods developed in Ref. \cite{BLLS} and Ref. \cite{paper1,paper2}. 
Those methods are both based on the following form of the Hamiltonian 
describing a discrete system $\{ q^A\}$ $(A=1,2,\cdots,n_{tot})$:
%
\begin{equation}
 	H_{0} = \frac{1}{2a}\delta^{AB}p_{A}p_{B} 
 		+ \frac{1}{2}V_{AB}q^Aq^B\ \ , 	\label{eqn:Hamiltonian}
\end{equation}
where $p_{A}$ is a momentum conjugate to $q^A$ and $a$ is a fundamental 
length of the system. For this discrete system it is easy to
divide the whole Hilbert space ${\cal F}$ into the form (\ref{eqn:F=F1*F2}): 
${\cal F}$ is defined as a Fock space constructed from $\{ q^A\}$
$(A=1,2,\cdots,n_{tot})$; ${\cal F}_1$ is defined as a Fock space
constructed from $\{ q^a\}$ $(a=1,2,\cdots,n_B)$; ${\cal F}_2$ is defined as
a Fock space constructed from $\{ q^{\alpha}\}$ 
$(\alpha=n_B+1,\cdots,n_{tot})$. In order to apply this scheme to our 
problem we have to construct a discretized theory of the scalar
field whose Hamiltonian is of the form (\ref{eqn:Hamiltonian}). 

First we expand the field $\phi$ in terms of the spherical harmonics as 
%
\begin{equation}
 \phi (\rho,\theta,\psi) =
 	\sum_{l,m}\frac{N^{1/2}}{r}\phi_{lm}(\rho)Z_{lm}(\theta,\psi )\ \ ,
\end{equation}
where $Z_{lm}=\sqrt{2}\Re Y_{lm}$ for $m>0$, $\sqrt{2}\Im Y_{lm}$ for 
$m<0$, and $Z_{l0}=Y_{l0}$. Then the Hamiltonian corresponding to the 
Killing energy for the action (\ref{eqn:action}) is decomposed into a 
direct sum of contributions from each harmonics component $H_{lm}$ as
%
\begin{equation}
 H = \sum_{lm}H_{lm}\ \ .\label{eqn:Hamiltonian2}
\end{equation}
Here $H_{lm}$ is given by
%
\begin{equation}
	H_{lm} = \frac{1}{2}\int d\rho\left[
		\pi_{lm}^2 + Nr^2
		\left\{\frac{\partial}{\partial\rho}\left(
		\frac{N^{1/2}}{r}\phi_{lm}\right)\right\}^2
		+ l(l+1)\left(\frac{N\phi_{lm}}{r}\right)^2\right]
		\ \ ,\label{eqn:subHamiltonian}
\end{equation}
where $\pi_{lm}(\rho )$ is a momentum conjugate to 
$\phi_{lm}(\rho )$.

Note that for any $(l,m)$ (\ref{eqn:subHamiltonian}) has the form  
%
\begin{equation}
	\bar{H} = \frac{1}{2}\int d\rho p^2(\rho)
		+ \frac{1}{2}\int d\rho d\rho'
		q(\rho)V(\rho,\rho')q(\rho')\ \ ,
		\label{eqn:form-of-Hamiltonian}
\end{equation}
where the following algebra of Poisson brackets is understood:
%
\begin{equation}
	\{q(\rho),p(\rho')\} = \delta(\rho -\rho') \ \ ,\ \ 
	\{q(\rho),q(\rho')\} = 0 \ \ ,\ \ 
	\{p(\rho),p(\rho')\} = 0 \ \ .
\end{equation}
Each subsystem described by the Hamiltonian (\ref{eqn:form-of-Hamiltonian}) 
can be discretized by the following procedure:
%
\begin{eqnarray}
	\rho & \to & (A-1/2)a\ \ ,\nonumber\\
	\delta(\rho -\rho') & \to & \delta_{AB}/a\ \ ,
	\label{eqn:discretization}
\end{eqnarray}
where $A,B=1,2,\cdots$ and $a$ is a cut-off length. The corresponding 
Hamiltonian of the discretized system is of the form (\ref{eqn:Hamiltonian}) 
with 
%
\begin{eqnarray}
	q(\rho) & \to & q^A\ \ ,\nonumber\\
	p(\rho) & \to & p_{A}/a\ \ ,\nonumber\\
	V(\rho,\rho') & \to & V_{AB}/a^2\ \ .
\end{eqnarray}
In this way we obtain a discretized system with the Hamiltonian 
(\ref{eqn:Hamiltonian}) with the matrix $V$ given by the direct sum 
%
\begin{equation}
	V = \mathop{\oplus}_{l,m}V^{(l,m)},
\end{equation}
where $V^{(l,m)}$ is explicitly expressed as
%
\begin{eqnarray}
 V^{(l,m)}_{AB}\phi_{lm}^A \phi_{lm}^B & = & 
	a\sum_{A=1}^{\infty}\left[
	N_{A+1/2}\left(\frac{x_{A+1/2}}{a}\right)^2\left(
	\frac{N_{A+1}^{1/2}}{x_{A+1}}\phi_{lm}^{A+1} -
	\frac{N_{A}^{1/2}}{x_{A}}\phi_{lm}^{A}\right)^2
	\right.	\nonumber\\
	& & \left.
	+ \frac{l(l+1)}{r_0^2}
	\left(\frac{N_{A}\phi_{lm}^{A}}{x_{A}}
	\right)^2		\right]\ \ .
\end{eqnarray}
Here
%
\begin{eqnarray}
	x_{A} & = & r(\rho =(A-1/2)a)/r_0\ \ ,\nonumber\\
	x_{A+1/2} & = & r(\rho =Aa)/r_0\ \ ,\nonumber\\
	N_{A} & = & N(\rho =(A-1/2)a)\ \ ,\nonumber\\
	N_{A+1/2} & = & N(\rho =Aa)\ \ ,\nonumber\\
	\phi_{lm}^{A} & = & \phi_{lm}(\rho =(A-1/2)a)\ \ .
\end{eqnarray}
In the matrix representation $V^{(l,m)}$ is given by the 
$n_{tot}\times n_{tot}$ matrix
%
\begin{eqnarray}
 \left( V^{(l,m)}_{AB}\right) & = & \frac{2a}{r_{0}^2}\left( 
               \begin{array}{cccccc}
                   \Sigma^{(l)}_1 & \Delta_1 & & & & \\
         \Delta_1 & \Sigma^{(l)}_2 & \Delta_2 & & & \\
       & \ddots & \ddots & \ddots & & \\
       & & \Delta_{A-1} & \Sigma^{(l)}_A & \Delta_A & \\
       & & & \ddots & \ddots & \ddots 
               \end{array}
             \right) \ \ \ , \nonumber\\
   \Sigma^{(l)}_A
          & = & \frac{1}{2}(r_{0}/a)^2N_{A}x_{A}^{-2}\left[
          	N_{A-1/2}x_{A-1/2}^2 + N_{A+1/2}x_{A+1/2}^2\right]
          	\nonumber\\
          	& & + \frac{1}{2}l(l+1)N_A^2x_A^{-2}\ \ ,\nonumber\\
   \Delta_A
          & = & -\frac{1}{2}(r_{0}/a)^2 N_{A}^{1/2}N_{A+1/2}
		N_{A+1}^{1/2}x_{A}^{-1}x_{A+1/2}^2 x_{A+1}^{-1}\ \ ,
		\label{eqn:Vlm}
\end{eqnarray}
where we have imposed the boundary condition 
$\phi_{lm}^{n_{tot}+1}=0$. 
In these expressions $r_0$ is an arbitrary constant, which we set to be 
the area radius $R_B$ of the boundary in the following arguments.

\section{Entanglement thermodynamics in Minkowski spacetime}

In the case of Minkowski spacetime $r(\rho)=\rho$ and $N(\rho)=1$. 
The discretized system $\{\phi_{lm}^A\}$ $(A=1,\cdots,n_{tot})$ is described 
by Hamiltonian of the form (\ref{eqn:Hamiltonian}) with the potential given 
by (\ref{eqn:Vlm}).

We split the system $\{\phi_{lm}^A\}$ $(A=1,\cdots,n_{tot})$ into the two 
subsystems, $\{\phi_{lm}^a\}$ $(a=1,\cdots,n_{B})$ and 
$\{\phi_{lm}^{\alpha}\}$ $(\alpha =n_{B}+1,\cdots,n_{tot})$. 
For this splitting, the entanglement entropy $S_{ent}$ and 
the entanglement energy $E_{ent}$ for the ground state can be 
calculated numerically. The results are expressed as 
%
\begin{eqnarray}
 S_{ent} & \propto & \frac{R_{B}^2}{a^2}\ \ ,\nonumber\\
 E_{ent} & \propto & \frac{R_{B}^2}{a^3}\ \ ,
\end{eqnarray}
for all definitions of $S_{ent}$ and $E_{ent}$, where $R_{B}$ is the radius 
of the boundary defined by $R_{B}=n_{B}a$~\cite{BLLS,paper1,paper2}. Hence 
the entanglement temperature $T_{ent}$ is proportional to $a^{-1}$: 
%
\begin{equation}
 T_{ent}\propto \frac{1}{a}\ \ .
\end{equation}
Thus the entanglement thermodynamics in Minkowski spacetime is quite 
different from the black-hole thermodynamics~\cite{paper1} since
$E_{BH}\propto R_{H}/l_{pl}^2$ and $T_{BH}\propto 1/R_{H}$.

\section{Entanglement thermodynamics in Schwarzschild spacetime}

In Schwarzschild spacetime the metric is given by
%
\begin{equation}
 ds^2 = -\left(1-\frac{R_{H}}{r}\right)dt^2 + 
 	\left(1-\frac{R_{H}}{r}\right)^{-1}dr^2 
			+ r^2 \left( d\theta + \sin^2{\theta}d\psi^2\right)\ \ ,
\end{equation}
where $R_{H}$ is the area radius of the horizon. As the radial 
coordinate $\rho$ we take the proper distance from the horizon:
%
\begin{equation}
	\rho = \frac{R_{H}}{2}\left[\sqrt{y^2-1} + 
		\ln\left( y+\sqrt{y^2-1}\right)
		\right]	\ \ ,\label{eqn:rho-r}
\end{equation}
where the variable $y$ is defined by $y=2r/R_{H}-1$. 

As in the case of the Minkowski spacetime we split the system 
$\{\phi_{lm}^A\}$ $(A=1,\cdots,n_{tot})$ into the 
two subsystems, $\{\phi_{lm}^a\}$ $(a=1,\cdots,n_{B})$ and 
$\{\phi_{lm}^{\alpha}\}$ $(\alpha =n_{B}+1,\cdots,n_{tot})$, and 
calculated the entanglement entropy and the entanglement energy $E_{ent}$ 
for the ground state numerically. Note that we consider only
degrees of freedom corresponding to positive $A$ (see Ref.~\cite{paper2} 
for the reason why we do not consider those corresponding to $A\le 0$). 

The results are expressed as
%
\begin{eqnarray}
 S_{ent} & \propto & \frac{R_{B}^2}{a^2}\ \ ,\nonumber\\
 E_{ent} & \propto & \frac{R_{B}}{a^2}\ \ ,\label{eqn:S&E_Sch}
\end{eqnarray}
for all definitions of $S_{ent}$ and $E_{ent}$, 
where $R_{B}$ is the area radius of the boundary defined by 
$R_{B}=r(\rho=n_{B}a)$ with $n_B=O(1)$~\cite{paper2}. Here note that the 
proportionality in (\ref{eqn:S&E_Sch}) holds for a fixed value of $n_{B}$. 
Hence the entanglement temperature $T_{ent}$ is proportional to
$R_{B}^{-1}$:  
%
\begin{equation}
 T_{ent}\propto \frac{1}{R_{B}}\ \ .
\end{equation}
Thus the entanglement thermodynamics in Schwarzschild spacetime has 
the same structure as that of the black-hole 
thermodynamics~\cite{paper2}.

\section{Entanglement thermodynamics in R-N spacetime}

In Reissner-Nordstr{\"o}m spacetime the metric is given by
%
\begin{equation}
 ds^2 = -\left( 1-\frac{2M}{r}+\frac{Q}{r^2}\right)dt^2 + 
		\left(1-\frac{2M}{r}+\frac{Q}{r^2}\right)^{-1}dr^2 
			+ r^2 \left( d\theta + \sin^2{\theta}d\psi^2\right)\ \ ,
\end{equation}
where $M$ and $Q$ are the mass and the charge of the black-hole. The area 
radii of the outer and the inner horizon $R_{H\pm}$ are 
%
\begin{equation}
 R_{H\pm} = M\pm\sqrt{M^2-Q^2}\ \ .
\end{equation}
As the radial coordinate $\rho$ we take the proper distance from the 
outer horizon
%
\begin{equation}
 \rho = \sqrt{(M^2-Q^2)(y^2-1)} + 
 	M\ln\left( y+\sqrt{y^2-1}\right)\ \ ,\label{eqn:rho-r'}
\end{equation}
where the variable $y$ is defined by 
%
\begin{equation}
 y = \frac{r-M}{\sqrt{M^2-Q^2}}\ \ .
\end{equation}

Now the numerical results for the entanglement entropy 
$S_{ent}$ and the entanglement energy $E_{ent}$ for the ground state 
are expressed as 
%
\begin{eqnarray}
 S_{ent} & \propto & \frac{R_{B}^2}{a^2}\ \ ,\nonumber\\
 E_{ent} & \propto & c(q)\frac{R_{B}}{a^2}\ \ ,\label{eqn:S&E_RN}
\end{eqnarray}
for all definitions of $S_{ent}$ and $E_{ent}$, where $R_{B}$ is the 
area radius of the boundary defined by $R_{B}=r(\rho=n_{B}a)$ with
$n_B=O(1)$, and the coefficient $c(q)$ is a function of
$q=Q/M$ given by 
%
\begin{equation}
 c(q) = \frac{\sqrt{1-q^2}}{1+\sqrt{1-q^2}}\ \ .
\end{equation}
The coefficient $c(q)$ approaches to zero in the limit $q\to 1$.
Hence the entanglement temperature $T_{ent}$ is
proportional to $R_{B}^{-1}$:  
%
\begin{equation}
 T_{ent} \propto \frac{c(q)}{R_{B}}\ \ .
\end{equation}
Note that $T_{ent}$ is zero in the extremal limit of the background
spacetime. Therefore as in the case of the Schwarzschild spacetime, the 
entanglement thermodynamics in the Reissner-Nordstr{\"o}m spacetime has 
the same structure as that of the black-hole thermodynamics.

\section{Conclusion}

In this paper we have constructed entanglement thermodynamics for a 
massless scalar field in Minkowski, Schwarzschild and 
Reissner-Nordstr{\"o}m spacetimes. The entanglement thermodynamics in Minkowski 
spacetime differs significantly from black-hole thermodynamics. On the 
contrary, the entanglement thermodynamics in Schwarzschild and 
Reissner-Nordstr{\"o}m spacetimes has the same structure as that of 
black-hole thermodynamics. In particular, it has been shown that entanglement 
temperature in the Reissner-Nordstr{\"o}m spacetime approaches zero in 
the extremal limit. 

Finally we comment on possible extensions of the entanglement
thermodynamics. The first is the inclusion of a charged field as matter. 
In particular, it will be valuable to analyze the entanglement 
thermodynamics for a charged field in Reissner-Nordstr{\"o}m
spacetime. In this case we can define the entanglement charge as an
expectation value of the charge of the field for the coarse-grained
state. This is now under investigation. The second is a generalization
to non-spherically-symmetric spacetimes. For example it is expected
that construction of entanglement thermodynamics in Kerr spacetime
requires the introduction of a concept of an entanglement 
angular-momentum.




\begin{thebibliography}{99}
\bibitem{Bekenstein}
J. D. Bekenstein, Phys. Rev. D{\bf 7}, 2333 (1973).
\bibitem{GSL}
S. Mukohyama, Phys. Rev. {\bf 56}, 2192 (1997); V. P. Frolov, D. N. Page, 
Phys. Rev. Lett. {\bf 71}, 3902 (1993). 
\bibitem{Hawking1975}
S. W. Hawking, Comm. math. Phys. {\bf 43}, 199 (1975).
\bibitem{BH-entropy}
E.g., V. Iyer and R. M. Wald, Phys. Rev. D{\bf 52}, 4430 (1995); 
V. P. Frolov, D. V. Fursaev and A. I. Zelnikov, 
Phys. Rev. D{\bf 54}, 2711 (1996);
M. Banados, C. Teitelboim and J. Zanelli, 
Phys. Rev. Lett. {\bf 72}, 957 (1994); 
J. D. Brown and J. W. York, Jr., Phys. Rev. D{\bf 47}, 1420 (1993); 
S. W. Hawking and G. T. Horowitz, Phys. Rev. D{\bf 51}, 4302 (1995); 
G. 't Hooft, Int. J. Mod. Phys. {\bf A11}, 4623 (1996); 
S. Mukohyama, Mod. Phys. Lett. {\bf A11}, 3035 (1996).
\bibitem{BLLS}
L. Bombelli, R. K. Loul, J. Lee and R. D. Sorkin, Phys. Rev. {\bf 
D34}, 373 (1986). 
\bibitem{Srednicki}
M. Srednicki, Phys. Rev. Lett. {\bf 71}, 666 (1993).
\bibitem{Sent}
For other recent discussions on  the entanglement entropy, 
see e.g., 
J. S. Dowker, Class. Quantum. Grav. {\bf 11}, L55 (1994);
D. Kabat, Nucl. Phys. {\bf B453}, 281 (1995); 
J. D. Bekenstein, gr-qc/9409015; 
S. Liberati, gr-qc/9601032; 
A. P. Balachandran, L. Chandra and A. Momen, hep-th/9512047; 
E. Benedict and S. Pi, Ann. Phys. {\bf 245}, 209 (1996); 
R. Muller and C. Lousto, {\bf D52}, 4512; 
C. Hozhey, F. Larsen and F. Wilczek, Nucl. Phys. {\bf B424}, 443
(1994); 
F. Larsen and F. Wilczek, Ann. Phys. {\bf 243}, 280 (1995); 
S. R. Das, Phys. Rev. {\bf D51}, 6901 (1995); 
S. Solodukhin, Phys. Rev. {\bf D54}, 3900(1996).
\bibitem{paper1}
S. Mukohyama, M. Seriu and H. Kodama, Phys. Rev. {\bf D55}, 7666.
(1997). 
\bibitem{paper2}
gr-qc/9712018, Thermodynamics of entanglement in Schwarzschild
spacetime, S. Mukohyama, M. Seriu and H. Kodama, submitted for
publication in Physical Review D. 



\end{thebibliography}
\end{document}